\def\G1915{GRS~$1915$+$105$}
\def\X1550{XTE~J$1550$--$564$}
\def\J1655{GRO~J$1655$--$40$}

\def\eg{{\it e.g.} }
\def\etal{et al. }
\def\ie{{\em i.e. } }
\def\correc#1{{\bf #1}}

\documentclass[]{aastex}
\usepackage{emulateapj5}
\usepackage{epsfig}
\usepackage{epsf}
\usepackage{color}
\definecolor{red}{rgb}{0.7,0,0}
\definecolor{blue}{rgb}{0,0,0.7}
\def\correc#1{}

\shortauthors{Rodriguez Corbel \& Tomsick}
\begin{document}
\shorttitle{XTE J1550--564 2000 Outburst : Spectral Analysis}
\title{Spectral evolution of the microquasar XTE J1550--564 over its entire 2000 
outburst}
\submitted{accepted for publication in ApJ}

\author{J. Rodriguez\altaffilmark{1,2}, S. Corbel\altaffilmark{1,3} and J.A. Tomsick\altaffilmark{4}}

\altaffiltext{1}{DSM/DAPNIA/Service d'Astrophysique (CNRS URA 2052), CEA Saclay, 91191 Gif-sur-Yvette, France}
\altaffiltext{2}{Integral Science Data Center, Chemin d'Ecogia, 16, CH-1290 Versoix, Switzerland}
\altaffiltext{3}{Universit\'e Paris VII Fédération APC, 2 place Jussieu, 75005 Paris, France.}
\altaffiltext{4}{Center for Astrophysics and Space Sciences, Code 0424, University of California at San Diego, La Jolla, CA 92093, USA}

\begin{abstract}
We report on RXTE observations of the microquasar \X1550 during 
a $\sim 70$ day outburst in April-June 2000. We present the PCA+HEXTE 
3--200 keV  energy spectra of the source, and study their evolution over 
the outburst. The spectra indicate that the source transited from an initial Low Hard 
State (LS), to an Intermediate State (IS) characterized 
by a $\sim 1$ Crab maximum in the 1.5--12 keV band, and then back to the LS. 
The source shows an hysteresis effect such that the second 
transition occurs at a 2--200 keV flux that is half of the flux at the first transition.
This behavior is similar to what observed in other sources and favors a common  
origin for the state transitions in soft X-ray transients. In addition, the first 
transition occurs at a $\sim$ constant 2--200 keV flux, which probably indicates a change 
in the relative importance of the emitting media, whereas the second transition occurs during
a time when the flux gradually decreases, which probably indicates that it is driven 
by a drop in the mass accretion rate.
 In both LS, the spectra are characterized by the presence of a strong 
power-law tail (Compton corona) with a variable high energy cut-off.
During the IS, the spectra show the presence of a $\sim 0.8$ keV thermal 
component which we attribute to an optically thick accretion 
disk. The inner disk radius as infered from disk-blackbody fits to the energy spectrum
remains relatively constant throughout the IS. This suggests that the disk may be 
close to its last stable orbit during this period. We discuss the apparently 
independent evolution of the two media, and show that right after
the X-ray maximum on MJD 51662, the decrease of the source luminosity is due to a 
decrease of the power-law luminosity, at a constant disk luminosity. The detection
of radio emission, with a spectrum typical of optically thin synchrotron emission, soon 
after the X-ray peak, and the sudden decrease of the power law luminosity 
at the same time may suggest that the corona is ejected and further detected as a discrete 
radio ejection. \\

\end{abstract}


\section{Introduction}
Soft X-ray transients (SXT) are accretion powered binary systems, 
hosting a compact object (either a neutron star or a black hole), which 
spend most of their life in quiescence, and are detected in X-rays as 
they undergo episodes of outburst. 
Their X-ray spectra are usually dominated by two components, representing 
different physical processes acting in the close vicinity of the accreting 
object. The soft X-rays are likely the spectral signature of an
 optically thick geometrically thin accretion disk, whereas the hard X-rays 
are interpreted as the inverse Compton scattering of the soft 
photons from the accretion disk on hot electrons present 
in an optically thin coronal medium. 
Depending on whether the electrons have a thermal velocity distribution or not,
 this ``hard tail'' can be characterized by the presence or absence of an 
exponential cut-off, at a given threshold energy. 
 Based on the shape and strength of the spectra one can distinguish between 
5 common spectral states thought to be linked to the accretion rate of
 the source (see \eg Belloni 2001 for a recent review).
\begin{itemize}
\item The {\itshape{Quiescent State}} (hereafter QS), is the ``off'' 
state in which SXTs spend most of their lives. The observations of SXTs
in such a state have shown that the spectrum was power law like, with a 
photon index that could be either
soft or hard (e.g. Kong et al. 2002). The luminosity is several orders 
of magnitude below that of the other states. The accretion disk is 
undetectable in the X-rays. 
\item In the {\itshape{Low Hard State}} (hereafter LS), 
the $\nu-f_\nu$ spectrum is peaked in the hard X-rays, and  characterized 
by a strong power law with a photon index $\Gamma \sim $ $1.5-1.9$, and a 
cut-off around 100 keV. 
The disk emission remains weak, and its innermost part has a temperature 
$kT \leq 0.5$ keV. 
\item During the {\itshape{Intermediate State}} (hereafter IS) the 
contribution 
of the two spectral components to the overall luminosity is of the same order. 
The disk reaches a temperature of $\sim 1$ keV, and the hard tail has a photon
 index
 of $\Gamma \sim 2.5$.
\item In the {\itshape {High Soft State}} (hereafter HS) the soft ($\leq 10$ 
keV) luminosity is high and the spectrum is dominated by a thermal component 
in that range. The temperature of the disk is $\sim 1-1.5$ keV. The power law 
is faint with a steep photon index $\Gamma \geq 2.5$.
\item In the {\itshape{Very High State}}, (hereafter VHS) the overall luminosity 
is close to the Eddington luminosity. The disk has a temperature $kT \sim1-2$ keV,
 and the power law is steep with $\Gamma \sim 2.5$, although both components 
contribute significantly to the luminosity. 
It should be noted that the VHS is spectrally intermediate between the LS and 
the HS (Rutledge et al. 1999), making then the IS and VHS similar states
observed at different luminosities (Homan \etal 2001, see also Méndez \& van 
der Klis, 1997).
\end{itemize}
\indent A summary of the source history can be found in Rodriguez et al. (2003, 
hereafter paper 1). \X1550 is a microquasar (Hannikainen et al 2001, Corbel et al. 2002)
hosting a black hole of $10.5 \pm 1.5$ $ M_\odot$ (Orosz \etal 2002), lying at a distance 
of $3.2$ kpc $\leq D \leq 10.8$ kpc, with a prefered distance of $5.3-5.9$ kpc 
(Orosz \etal 2002). Extensive spectral and timing analysis of its 1998-1999 
outburst has shown  the need of an additional parameter beside the accretion 
rate $\dot{M}$ to account for the X-ray state transitions (Homan et al. 2001). \\
\indent Renewed X-ray activity of \X1550 was reported by Smith \etal (2000). 
The source underwent a $\sim 70$ day outburst starting on 2000 April 6
(paper 1, and reference therein).  Corbel \etal (2001) report 
the detection of 
 a radio emission with a negative spectral index on MJD 51665. They attribute
 it to optically thin synchrotron emission from a discrete 
ejection. The date of the ejection event is however hard to constrain, and may
 correspond to the state transition occuring a few days before 
(Corbel \etal 2001). These authors also point out the absence of radio emission
 on MJD 51670, indicating that no jet feature is present during that 
time, and they detect radio emission with an inverted spectrum on MJD 51697, which 
they attribute to a compact jet. The outburst initiates
in the IR and optical (Jain \etal 2001) $\sim 10$ days before 
the X-rays, and a second IR/Optical maximum occurs as the source has 
returned to the LS around MJD 51690. 
This second IR-Optical peak is possibly related to the compact jet  
synchrotron tail (Corbel \etal 2001) as its inverted spectrum 
can extend up to the near IR range (Corbel \& Fender, 2002). 
Tomsick, Corbel \& Kaaret (2001) (hereafter TCK01) report spectral 
(RXTE+Chandra) observations during the 
very last part of the outburst, as \X1550 is returning to quiescence.\\ 
\indent We have studied the behavior of a low frequency QPO in paper 1, 
and we focus here on the 
X-ray spectral behavior of \X1550 from the very beginning of 
the PCA+HEXTE pointed observations on MJD 51644, until MJD 51698 where the 
observations are contaminated both by the Galactic ridge diffuse emission 
and the close outbursting transient pulsar XTE J$1543-568$ (TCK01). We perform 
an analysis similar to that reported in TCK01 (covering MJD 51680--51698), and 
add the whole RXTE data set publicly available in the archives covering this outburst. 
We thus present for the first time the entire PCA+HEXTE spectral analysis of 
\X1550 over its 2000 outburst. 
The organization of the paper is as follows; we 
start by presenting the data reduction and analysis method used, before 
presenting the spectral evolution of the source. We discuss our results 
in the last part of the paper.

\section{Observations}
\subsection{Data Reduction}
\label{sec:analysis}
\X1550 has been observed continuously with RXTE all over its outburst, \ie 
from MJD 51644 (Apr. 10), to MJD 51741 (July 16). 
Since the background in the Galactic ridge is difficult to estimate with a 
non-imaging instrument, we restrict our study to the interval between MJD 
51644 and MJD 51698, and refer the reader to the study of TCK01 for the 
following period.\\
\indent We have reduced and analyzed the data using 
{\itshape {LHEASOFT}} package v5.2, which includes new response matrices  
and new background files for the PCA. We use here the maximum number of PCUs 
turned on over an observation, and both clusters of HEXTE.
We restricted ourselves to the time when the elevation angle was above  
$10^{\circ }$, the offset pointing less than $0.02^{\circ }$, and we 
also rejected  the data taken while crossing the SAA. 
In addition, the good time intervals (GTIs) were defined when the number 
of Proportional Counter Units (hereafter PCU) turned on was constant 
and equal to the maximum available over a given observation, for most of them,
at least 3 PCUs were turned on.  May 12, 2000 
corresponds to the abrupt loss of Xenon layer in PCU 0, 
which renders its use for spectral analysis difficult and uncertain.
We therefore extracted all the spectra from the top layer of all available 
PCUs, except PCU 0. All spectra from individual PCUs were sumed during the extraction. 
Background spectra were estimated using {\em{PCABCAKEST V.3.0}}.
The responses were generated with {\em{pcarsp V8.0}}.\\ 
\indent HEXTE spectra were extracted from both clusters,
 in the standard mode data. 
 We followed the ``cook book'' procedures for separating the ON and OFF 
positions before extracting the raw spectra (source and background), 
and then correct them for dead-time. Responses were estimated with {
\em {hxtrsp V3.1}}. The PCA+HEXTE resultant spectra of a single observation 
were then analyzed together in {\em {XSPEC v11.1.0}}. 
We retain in our fits the energy channels between $3$ and $30$ keV for the PCA,
 and between $18$ and $200$ keV for the HEXTE. Furthermore, in order to 
accommodate for uncertainties in the PCA response matrix, we 
 included $0.8\%$ systematic errors from 3--8 keV, and $0.4\%$ from 8-30 keV
(see TCK01).\\

\subsection{Spectral analysis}
\begin{figure}
\plotone{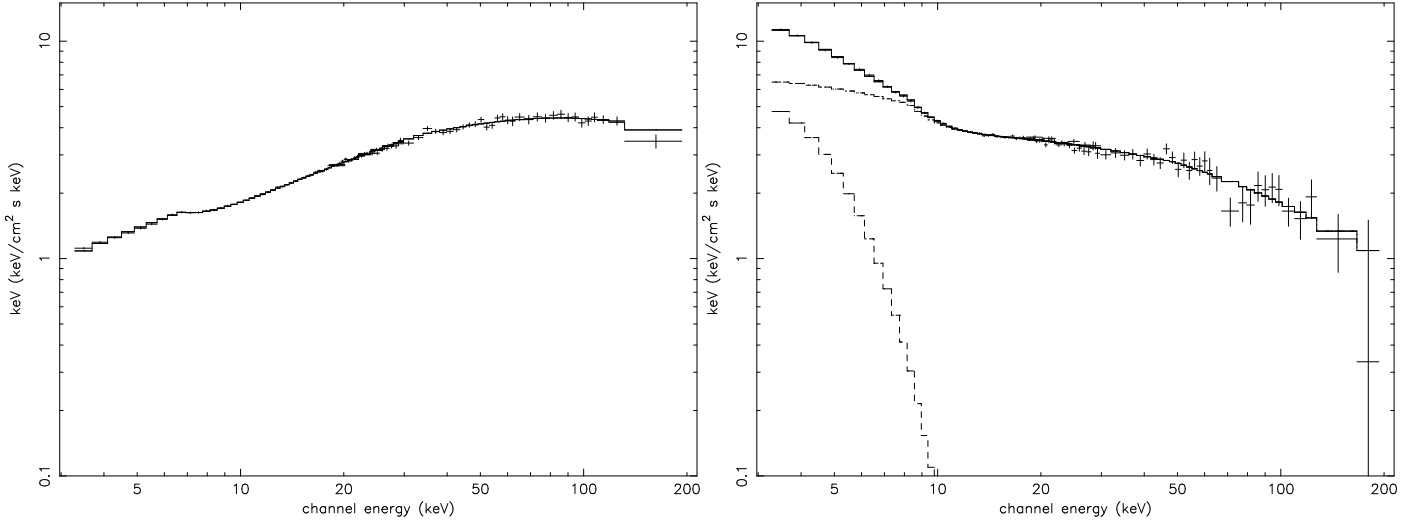}
\caption{PCA ($3-30$ keV) and HEXTE ($18-200$ keV) unfolded spectra. 
{\em{Left}} shows a typical LS spectrum, with the 
 best fit model (see the text for the details of the modeling) over-plotted 
(line). 
{\em{Right}} is a IS spectrum showing the 
different spectral components (dashed) used in the fits, {\it i.e.} a 
multicolor disk black body plus a power law. The best fit model is 
over-plotted as a line. The spectra are in ``$\nu F_{\nu}$'' units.}
\label{fig:spectra}
\end{figure}

Several models were tested in the course of the spectral analysis. 
In every fit, a multiplicative constant, 
representing  the normalization between the instruments, was added to
 the spectral model in order to take into account the uncertainties in the 
PCA-HEXTE cross calibration. 
To  determine the spectral models, we first fitted PCA+HEXTE from  
MJD 51646 and MJD 51648 with a simple model consisting of 
interstellar absorption ({\em{wabs}} in XSPEC terminology) plus a power law. 
The resultant $\chi^2$ is
 poor (1191 for 142 dof), with large residuals around 6.5 keV and a broad 
minimum
 around 10 keV. As the RXTE bandpass is not ideally suited for the determination
 of the interstellar absorption, we applied the equivalent Hydrogen column 
density value $N_H$ returned from recent CHANDRA observations (Kaaret et al. 
2003), \ie 
$N_H=0.9\times 10^{22}$ cm$^{-2}$, and froze it in all our fits. 
Adding iron edge absorption ({\em{smedge}}) improves the fits significantly 
 with $\chi^2=942$ (139 dof). Following  Sobczack \etal (1999), and TCK01, 
we froze the width of the smedge model to 10 keV.
In order to accommodate for the high energy behavior (HEXTE band) an 
exponential cut-off at higher energy is needed and  gives satisfactory fits 
with $\chi^2=137$ (138 dof) for MJD 51646, and $144$ (138 dof) for MJD 
51648.
  Adding a Gaussian emission line around 6.5 keV gives only a marginal 
 improvement to our fits. We also tentatively included a 
 multi-color disk blackbody (Mitsuda \etal 1984), thought to represent the 
emission coming from the optically thick accretion disk, but our fits failed to
 achieve convergence, at least from MJD 51644 through MJD 51655. 
The spectral parameters returned from the fits are shown for the 
entire outburst in Table 
\ref{tab:specparam}, and the spectrum of MJD 51644 is represented in the 
left panel of Fig. \ref{fig:spectra}.\\
\indent From  MJD 51658 a soft excess is detectable in the spectra  
(Fig. \ref{fig:spectra} shows the example of  MJD 51662). 
Using the model consisting of an absorbed (Hydrogen + Fe Edge)
 power law cut at high energy leads to poor $\chi^2$ ($153$ for 
$138$ dof, and even worse for the following days). The addition of a 
multi-color disk blackbody model (Mitsuda \etal 1984) to the fits 
greatly improves their quality 
($\chi^2=134$ for $136$ dof on MJD 51658, although the disk radius 
parameter is difficult to constrain that day). Note that during IS/VHS, 
the multi-color disk model can give unreliable values of the inner 
radius (Sobczak et al. 1999, Merloni, Fabian \& Ross 1999), especially when the 
ratio of the disk flux to the total flux is low (Merloni et al. 1999).
The high energy 
cut-off is still needed until MJD 51662 (it is significant at the 99.5 $\%$ 
level using an F-test). We have checked its 
presence the following days by systematically adding an exponential cut 
off in our fits, but it gave no improvement to the fits (see Table 
\ref{tab:specparam}). Over that period, the hard X-ray contribution 
has significantly decreased 
(Fig. \ref{fig:hardstate}), and the photon index ranges
 from $\simeq 2.0$ to $\simeq 2.4$. The iron line may still be present but
 for the same reasons as explained above we did not include it in our analysis.
\\
\indent Around MJD 51683  the fits fail to converge if the 
blackbody component is kept in the model. 
From MJD 51682 until MJD 51688, an exponential cut-off is needed in the fits. 
We note, however, that for 
MJD 51686, 51687, and 51688, the resultant improvement in the fits becomes 
only marginal; we nonetheless kept this additional component. 
After MJD 51688 there is no convergence  when the exponential cut-off 
is included in the model (see also TCK01). 
So that the last observations are fitted with a simple model consisting 
of interstellar and smeared edge ($\geq 7$ keV) absorptions, and a power law.\\
\indent Two of the observations were already reported by Miller et al. 
(2003, MJD 51667, 51670). Although they included
a Gaussian feature in their fit, we note the relatively good agreement between their results and 
ours for both dates. The small differences can be due to the fact that they have used a  
value of $0.8\times 10^{22}$ 
cm$^{-2}$ for the N$_H$ parameter, and have made joint Chandra+RXTE spectral fits.\\

\begin{table*}[thpb]
\footnotesize{
\begin{tabular}{|ccccccc|}
Date   & Photon Index & Color Radius$^*$ & Disk Temp. & $E_{cut-off}$ & $E_{fold}$ & $Red. \chi^2$ \\
 (MJD)    &          &    (km)     &   (keV)    &    (keV)      &    (keV)   &  (d.o.f.)\\
\hline
51644.4 & $1.49\pm0.01$ &  &   & $33.8_{-2.1}^{+2.0}$ & $165 \pm7$ & 1.10 (138)\\
51646.3 & $1.46\pm0.01$ &    &     & $33.6_{-3.0}^{+2.8}$ & $137_{-7}^{+8}$ & 0.99 (138)\\
51646.6 & $1.48\pm0.01$ &    &     & $35.9_{-2.5}^{+2.3}$ & $127_{-5}^{+6}$ & 1.26 (138)\\
51648.7 & $1.47\pm0.01$ &    &     & $33.3\pm1.6$ & $123_{-3}^{+4}$ & 1.04 (138)\\
51650.7 & $1.47\pm0.01$ &    &     & $30.3\pm1.8$ & $120\pm4$ & 1.24 (138)\\
51651.4 & $1.48\pm0.01$ &    &     & $31.9_{-1.5}^{+1.7}$ & $122_{-4}^{+3}$ & 1.34 (138)\\
51652.2 & $1.47\pm0.01$ &    &     & $33.6_{-2.0}^{+1.6}$ & $111\pm4$ & 1.11 (138)\\
51653.5 & $1.50\pm0.01$ &    &     & $32.1_{-1.3}^{+1.2}$ & $109\pm3$ & 1.00 (138)\\
51654.7 & $1.51\pm0.01$ &    &     & $29.7\pm1.5$ & $109\pm3$ & 0.96 (138)\\
51655.7 & $1.55\pm0.01$ &    &     & $30.7_{-2.1}^{1.7}$ & $111\pm5$ & 0.97 (138)\\
51658.6 & $1.70\pm0.01$ & $146_{-108}^{+1413}$ & $0.37\pm0.014$    & $19.4_{-1.5}^{+1.6}$ & $115\pm5.6$ &0.98 (136)\\
51660.0 & $2.05\pm0.02$ &$47_{-12}^{+19}$ & $0.60\pm0.06$ & $18.7_{-0.7}^{+1.1}$ & $144_{-7}^{+8}$ & 0.92 (136)\\
51662.2 & $2.33\pm0.02$ & $32.5_{-1.3}^{+2.5}$ & $0.95\pm0.04$ & $17.7_{-2.0}^{+4.3}$  & $422_{-80}^{+216}$  & 0.86 (136)\\
51664.4 & $2.37\pm0.02$ & $45.6_{-1.5}^{+0.9}$ & $0.86\pm 0.01$ &   &   & 1.05 (138)\\
51664.6 & $2.35\pm0.02$ & $42.4\pm1.5$ & $0.89\pm0.01$ & & & 1.43 (138)\\
51665.4 & $2.38\pm0.02$ & $52.0_{-1.9}^{+2.7}$ & $0.80\pm 0.01$ &   &   & 1.00 (138)\\
51667.7 & $2.28\pm0.02$ & $47.8_{-1.3}^{+2.4}$ & $0.82\pm0.02$ &  &  & 0.95 (138)\\
51668.8 & $2.28\pm0.02$ & $47.8_{-1.3}^{+1.4}$ & $0.80\pm0.01$ &  &  & 1.07 (138)\\
51669.2 & $2.30\pm0.01$ & $49.5_{-1.7}^{+1.9}$ & $0.78\pm 0.01$ &  &  & 1.18 (138)\\
51670.5 & $2.28\pm0.01$ & $53.0_{-1.5}^{+2.2}$ & $0.77\pm0.02$ &  &  & 1.02 (138)\\
51670.8 & $2.26\pm0.02$ & $ 49.7_{-2.3}^{+1.9}$ & $0.78\pm0.01$ & &  & 0.87 (138)\\
51671.4 & $2.30\pm 0.02$ & $50.8_{-2.0}^{+1.8}$ & $0.75\pm0.01$ &  &  & 1.06 (138)\\
51672.4 & $2.27\pm 0.02$ & $55.2\pm2.2$ & $0.75\pm 0.02$ &  &  & 1.08 (138)\\
51672.9 & $2.32\pm0.01$ & $51.0_{-1.7}^{+3.0}$ &$0.77\pm0.02$ & & & 1.20 (138)\\
51673.4 & $2.31\pm0.01$ & $39.4_{-2.0}^{+2.5}$ & $0.77\pm0.02$ &  &  & 1.28 (138)\\
51674.7& $2.16\pm 0.02$ & $45.0\pm2.4$ & $0.77\pm 0.02$ &  &  & 1.52 (133)\\
51675.4 & $2.22\pm 0.01$ & $43.8\pm1.8$ & $0.67\pm 0.02$ &  &  & 0.92 (138)\\
51676.4 & $2.18\pm0.01$ & $46.1_{-4.9}^{+5.1}$ & $0.65\pm0.03$ &  &  & 1.16 (138)\\
51678.5 & $2.06\pm0.01$ & $46.5_{-4.7}^{+6.3}$ & $0.54\pm0.05$ &  &  & 1.23 (138)\\
51680.4& $2.01\pm0.01$ & $53.5_{-14.0}^{+17.8}$ & $0.56\pm0.06$ &  &  & 1.16 (138)\\
51682.3 & $1.76\pm0.01$ & $39_{-12}^{+20}$ & $0.5\pm0.1$ & $30.3_{-4.9}^{+6.2}$ & $198_{-32}^{+37}$ & 0.93 (136)\\
51683.8 & $1.638\pm0.005$ &    &     & $60.5_{-6.6}^{+7.4}$ & $197_{-31}^{+38}$ & 0.84 (138)\\
51684.8 & $1.647\pm0.007$ &    &     & $59.2_{-9.9}^{+23.1}$ & $185_{-55}^{+63}$ & 0.91 (138)\\
51686.3 & $1.60\pm0.01$ &    &     & $76.4_{-11.6}^{+12.6}$ & $302_{-108}^{+257}$ & 1.00 (138)\\
51687.2 & $1.53\pm 0.01$ &    &     & $55.8.4_{-11.5}^{+11.0}$ & $202_{-52}^{+78}$ & 1.17 (138)\\
51688.8 & $1.49\pm 0.01$ &    &     & $31.8_{-17.8}^{+11.7}$ & $338_{-65}^{+114}$ & 0.78 (138)\\
51690.7 & $1.48\pm0.02$ &    &     &  &  & 1.05 (140)\\
51690.9 & $1.53\pm0.01$ &  & & & & 1.32(140)\\
51692.5 & $1.48\pm0.01$ &    &  &  &  & 1.17 (140)\\
51693.4 & $1.50\pm0.02$ &    &     &  &  & 1.07 (140)\\
51695.2 & $1.50\pm0.02$ &    &     & & & 0.87 (140)\\
51696.4 & $1.53\pm0.01$ &    &     & &  & 1.04 (140)\\
51698.9 & $1.59\pm0.05$ &    &     &  &  & 0.90 (140)\\
\hline
\end{tabular}}
\caption{Best fit parameters returned from the fits over the whole outburst. \newline $^* R_{col}/1km=\sqrt{\frac{Norm}{\cos i}} \times \frac{D}{6 kpc}$, where $i=73.1^\circ$ (see Orosz \etal 2002).}
\label{tab:specparam}
\end{table*}

\section{Spectral evolution : global behavior and state identification}
The evolution of the spectral parameters are reported in Table 
\ref{tab:specparam} and  they are plotted on Fig. \ref{fig:hardstate}. 
The source spectral behavior over the outburst can be divided into two 
 distinct states as illustrated by Fig. \ref{fig:hardstate}. 
Based on the 
spectral parameters returned from the fits (Table \ref{tab:specparam}), 
we identify, without any ambiguities, the initial
 rise as a standard LS (from  MJD 51644 to  MJD 51658). 
The evolution of the source is rather slow here; the photon 
index rises from $1.49\pm0.01$ to $1.70\pm0.01$ (Fig. \ref{fig:hardstate}), 
as the $2-60$ keV flux slowly rises 
from $1.3 \times 10^{-8}$ erg/cm$^2$/s (MJD 51644) to $2.0\times 10^{-8}$ 
erg/cm$^2$/s (MJD 51658). The exponential cut-off ranges from $\sim20$ to 
$\sim36$ keV, and presents no obvious 
correlation with the flux, whereas the folding energy decreases 
from $\simeq 165$ keV, on MJD 51644 to $\simeq 115$ keV 
on MJD 51658, with an increasing soft flux. MJD 51658 is still 
typical of a LS, with a photon index $<2$, but here the spectrum has already 
started to soften (Fig. \ref{fig:hardstate}).\\
\indent The state transition occurs on MJD 51660, where the hard tail
 steepens significantly with a photon index $\geq 2$. On MJD 51660 and 51662, 
however, a cut-off at high energy is still needed in the fits. We note that 
the folding energy increases up to $\simeq 144$ keV on MJD 51660 and $\simeq 422$
keV on MJD 51662. Based on the spectral parameters, we can identify this new
 state as an intermediate/very high state.
It is difficult to distinguish between these two states since both have similar
spectral and temporal behaviors (and may be the same state observed at 
different luminosities, Homan \etal 2001, M\'endez \& van der Klis 1997). 
The presence of high frequency QPOs (Miller \etal 2001) over the period of 
IS/VHS, does not help much in lifting the ambiguity. 
It is usually assumed that the VHS has a luminosity close to $L_{Edd}$, which 
in our case, assuming a $10M_\odot$ black hole at $6$ kpc (although the 
uncertainties are large), is $\sim 1.3\times 10^{39}$ erg/s. 
Here at the maximum (MJD 51662), the bolometric disk flux is
$2.1 \times 10^{-8}$ erg/cm$^2$/s, and the 2--500 keV (extrapolated) power-law
flux is $3.35 \times 10^{-8}$ erg/cm$^2$/s, giving a bolometric luminosity 
close to $\sim 2.3 \times 10^{38}$ erg/s. Given the flux and in 
comparison with the previous outburst (Sobczak \etal 1999), we will refer to 
this state as an IS.\\
\indent The spectral evolution from MJD 51660 to MJD 51662, is rather abrupt 
(Fig. \ref{fig:hardstate}), and  
suggests that the source has undergone dramatic evolution. Indeed, the power 
 law tail is now much steeper, the disk temperature higher, and the folding energy
of the cut-off has increased by a factor of $\sim 3$. 
 The (color) radius obtained from the spectral fits (Fig. 
\ref{fig:hardstate}) varies 
between $\simeq 32$ to $\simeq 146$ km (assuming a distance of $6$ kpc, for 
$73.1^\circ$ inclination). Its value remains constant around  
$45-55$ km over 13 observations.  
The disk reaches its highest temperature on MJD 51662, and
from MJD 51665 until MJD 51674  its color temperature 
is around $0.75-0.8$ keV and is found fairly constant 
(Fig. \ref{fig:hardstate}). 
After that, it starts to decrease down to $\sim 0.5$ keV on MJD 51682. We note 
here that, although the color temperature decreases, the color radius seems 
to decrease also. This may be due to the difficulty in determining this
parameter from spectral fits starting above 3 keV at times where the source
count rate starts to be low (compared to e.g. MJD 51658, where although the 
disk has a lower color temperature, the counting statistics allow a better 
estimate of the parameters).\\
\indent After MJD 51680, the source slowly returns into a LS 
(see TCK01 for the spectral analysis of the observations after MJD 51680). 
Although the transition to the final low state is not as sharp as in 
the initial  stage, there is a clear evolution in terms of spectral 
parameters, between MJD 51680 and MJD 51682, where the spectrum gets harder
(Fig. \ref{fig:hardstate}), 
and manifests a cut-off at high energies (Table \ref{tab:specparam}). 
From MJD 51682 to 51698 the source is in a LS, which first shows 
the presence of an exponential cut-off at high energy (MJD 51682-51688),
while the following observations do not show any  cut-off up to 200 keV.\\
\indent Our timing analysis (paper 1) further confirms the nature of the 
states. During both states, we detect low frequency QPOs in the range 0.1--10 
Hz. Two types of LFQPOs seem to be present over the outburst. The first type 
resembles the type C QPO (Remillard \etal 2002), and the other is more likely 
a type A. Interestingly, the presence of type C QPO is not related to the 
spectral state of the source, since it disappears after the X-ray peak on MJD 51662 
(Fig. \ref{fig:hardstate}), well after the first transition, 
and reappears after the secondary peak, on MJD 51674, before the second transition.
 During the LS, however, it has a high amplitude (10--16 $\%$ RMS), and during 
the IS it is fainter (5$\%$) as usually observed. 

\section{Discussion}
\subsection{Hard Component Evolution During the Initial Hard State}
Although  recent studies have shown that the high energy emission could 
originate in a compact jet  (Markoff, Falcke \& Fender, 2001; 
Markoff \etal 2003), the results we present here will be discussed in 
the context of the standard Comptonization model involving a corona.
We note, however that the coronal geometry is unclear, and the corona could 
be seen as the base of the jet. \\
\indent In the standard Comptonization picture, the folding energy of the cut-off 
is close to the electron temperature. 
 The slight decrease of the folding energy between
MJD 51644--51658 may indicate that the Compton cloud is cooled more 
efficiently. This might correspond to the approach of the accretion disk
(decreasing inner radius), while its inner temperature and the soft X-ray flux
 increase. This interpretation is in good agreement with the state 
transition observed on MJD 51660, and the presence of a 
thermal component in our fits from MJD 51658. This is also in good agreement 
with the 
observation of an infrared-optical luminosity peak with a spectrum that is 
compatible with a thermal emission, occuring
 10 day before that in the X-ray (Jain \etal 2001). In that case, the 
10 day delay between the infrared/optical and the X-rays would indicate
 that the accretion disk fills on a viscous time scale suggesting that it is 
truncated at a certain distance from the black hole.\\
\indent The fact that the photon index evolves from $\Gamma=1.57$ on MJD 51655 to 
$\Gamma=1.74$ on MJD 51658 (Fig. \ref{fig:hardstate}) may
indicate that the state transition initiates first by a change in the 
Compton medium. As already suggested from the analysis of the Unconventional Stellar Aspect 
(USA) data (Reilly \etal 2001) the situation could be similar to the two component 
accretion flow proposed in the case of other sources  
(Smith, Heindl \& Swank, 2002). In that case, both flows possess their own 
response time-scale to any external perturbation (of the accretion rate). 
In a standard thin disk it corresponds to the viscous time-scale 
$t_{visc} \sim \frac{R^2}{\nu}$ 
(R being the radial distance to the black hole, and $\nu$ the kinematic 
viscosity, usually parameterized with the $\alpha$ prescription), 
which is then of the order of days (Frank, King \& Raine, 1992). 
The coronal time-scale is the free fall time-scale, which is much shorter. 
In this picture, any external change in the accretion rate would first have an 
incidence on the corona, and then on the disk. We should note here that the 
evolution of \X1550 during this LS is very similar to that of the 1998 
outburst (Wilson \& Done, 2001; Wu et al. 2002). This suggests a common evolution 
for both outbursts although \X1550 did not reach the same X-ray luminosity during the 
2000 outburst.

\subsection{State Transitions and Hysteresis Effect Between the Two LS}
The first transition occurs at a total 2-200 keV flux of $\sim 2.3 \times 10^{-8}$ 
erg/cm$^2$/s on MJD 51660, and the second on MJD 51682 with a total 2-200 keV flux of 
$1.1 \times 10^{-8}$ erg/cm$^2$/s. The first transition 
is rather abrupt and occurs at a
roughly constant flux, whereas the second is smoother and occurs while the source 
flux is monotonically decreasing, at a flux half that of the first one.  
We have plotted on Fig. \ref{fig:alphavsflux} (left panel) the evolution
 of the spectral index vs. the 2--200 keV flux over the outburst. The figure shows an 
hysteresis, similar
 to what is observed in other black hole sources (Miyamoto \etal 1995, 
Nowak \etal 2002), and possibly in the neutron star Aql X--1 (Maccarone \& Coppi 2003). 
In order to have a direct comparison with the latter source, we have
plotted on Fig \ref{fig:alphavsflux} (right panel) the ratio of the 2--50 keV 
powerlaw (corona) flux to the 2--50 keV disk-blackbody flux versus the 2--50 keV source 
flux. The behavior of XTE~J1550--564 is here similar to that of Aql X--1 shown on Fig. 1 
of Maccarone \& Coppi (2003). In addition, Maccarone \& Coppi (2003) have shown
that the evolution of the hardness ratio versus the soft (1--12 keV) X-ray flux, 
i.e. the evolution of a given  source in terms of spectral states versus source flux, 
was very similar for at least 4 black hole SXTs (XTE J1550--564, XTE J1859+326, 
XTE J2012+381, and probably XTE J1748--288, in addition to those previously
reported by Miyamoto \etal 1995, i.e. GX 339--4, GS 1124--683 and Cyg X--1, and 
Nowak \etal 2002, GX 339--4), with the occurence of a  
hysteresis loop. This similarity, both with black hole systems, and at least one 
neutron star system suggests that the state transitions, and the spectral evolution of those 
sources over their outbursts, obey similar mechanisms.\\ 
\indent The fact that the first transition occurs at constant flux is not what
is expected if we assume that the increase of the  luminosity 
is  due to an increase of the accretion rate alone. This behavior more likely 
reflects a change in the relative importance of the emitting media as pointed out 
by Zhang \etal (1997) in the case of Cyg X--1. This
 would further confirm  the need of an additional parameter (beside the 
accretion rate) to model  the state evolutions of LMXBs in outburst as already 
pointed out by Homan \etal (2001) during the previous outburst of \X1550, and 
Smith \etal (2002) for other sources.  The additional parameter
beside the accretion rate may then be related to the relative geometry disk-corona 
(e.g. the corona size as suggested by Homan \etal 2001). In that case a 
stable accretion rate will produce the same amount of energy, and the relative geometry 
of the disk-corona system will modify the spectral energy distribution.\\
\indent During the decline of the outburst, the transition towards the LS occurs 
while the flux is dropping similarly in all spectral bands (Fig. \ref{fig:hardstate}). 
This behavior is what expected if the state transition is driven only by 
variations of the accretion rate.
In addition, when comparing the two different LS, one can see that the folding energy of the
cut-off is systematically higher during the second LS (when present). This indicates 
that the coronal temperature is higher during the second LS. If we assume that 
the electron temperature depends on the flux of soft photons from the disk, the
higher folding energy after the outburst may indicate that the disk recedes faster than 
it approaches the black hole. Although our observations seems to indicate rather constant
physical parameters for the disk (Fig. \ref{fig:hardstate}), the fact that the disk is 
suddenly not detected soon after the transition, at times where the cut-off re-appear in 
the spectra, provides evidence for this possibility. The recession of the disk is 
compatible with the observation of 
a 65 Hz QPO by Kalemci \etal (2001), which seems to be of the same type as the
HFQPO previously seen during the brightest part of the outburst (Miller et al.
 2001).\\

\subsection{The Intermediate State: Accretion Disk Behavior}
The presence of a cut-off in the spectra of MJD 51660 and MJD 51662 
suggests that the coronal electrons have still a thermalized velocity 
distribution, although the photon index is soft. Those days also, the disk 
starts contributing significantly to the overall luminosity. So, in the 
standard picture of Comptonization, as the amount of cool thermal photons 
increases, the coronal electron temperature should decrease due to the 
higher cooling rate.  The observation of an increasing folding energy
for the cut-off between MJD 51658 and MJD 51662 argues, on the contrary, for coronal 
heating. This behavior is similar to what is observed by TCK01 although during 
the decay. They suggest that this might be the signature of the onset of different emission 
mechanism (e.g. bulk motion Comptonization).\\
\indent According to  Merloni et al. (1999), low values 
of the disk radius and flux ratio 
 hide some failures in the basic multi-color disk-blackbody model. In a 
previous work (Rodriguez \etal 2002a), we have retained the radius values 
in our analysis only 
when $\frac{F_{bbody}}{F_{tot}}>0.5$, with the fluxes estimated between 
2--50 keV. This criterion is more stringent than that of Merloni \etal 
(1999), and has allowed us to study the disk radius behavior with high 
accuracy (Rodriguez \etal 2002a). On MJD 51662, the $\sim2-50$ keV 
disk unabsorbed flux is $8.4\times 10^{-9}$ erg/cm$^2$/s, giving a  
$\sim 25\%$ contribution to the total flux. It is very likely, then, that 
the disk parameters returned from the fit this day are unreliable. The disk
behavior over the 2000 outburst resembles that of the 1998 outburst (at least during 
the first part of the latter reported in Sobczak et al. 1999). Although the peak 
temperatures, and the luminosities of the two outbursts differ significantly, there are, 
in both cases periods of relatively constant disk parameters. During these, the temperatures 
are in the same ranges (0.5--0.9 keV, Table 1 in Sobczack et al. 1999, 
Table \ref{tab:specparam} in the current study). Given the  
constancy of the disk parameters, especially the inner radius over the IS 
(Fig. \ref{fig:hardstate}), and the similarity with the huge 
1998-1999 outburst, it is tempting to consider that the disk is close to the last stable 
orbit. The presence, over this period, of high frequency QPOs ($251-276$ Hz, 
Miller \etal 2001), almost the highest values as yet observed in this source
  ($285$ Hz, Remillard \etal 1999, Homan \etal 2001), may further support 
this assumption (see also Kalemci et al. 2001). \\

\subsection{Ejection of the corona ?}
\begin{figure}
\plotone{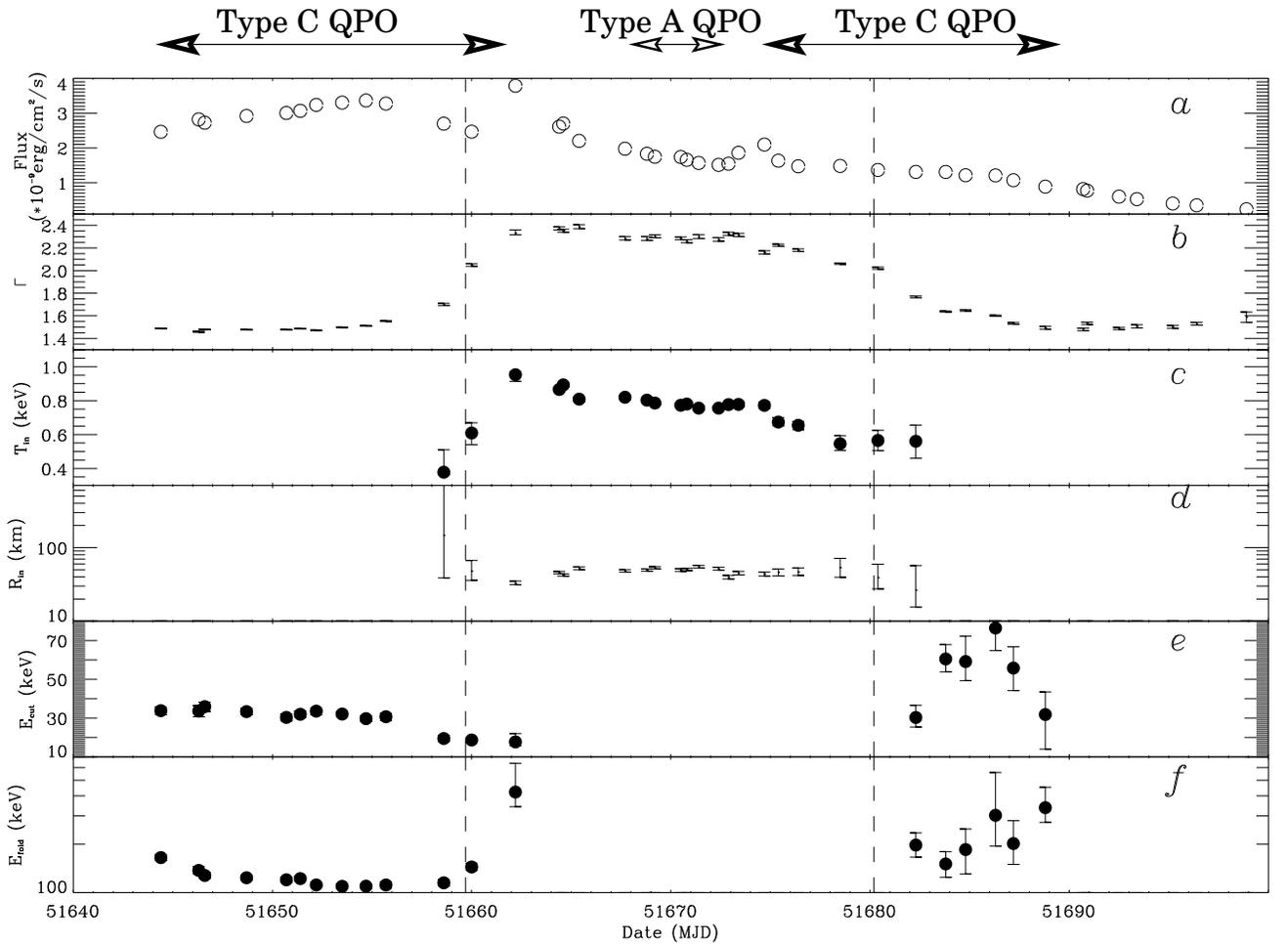}
\caption{Evolution of the spectral parameters over the outburst. $a)$  2--200 keV 
 light curve of the outburst.
$b)$ Power law spectral index $\Gamma$. $c)$ Inner disk  temperature (keV). 
$d)$ Inner disk  radius (assuming D=6 kpc and $i=73.1^\circ$).  $e)$ cut-off 
energy (keV). $f)$ e-folding of the cut-off (keV). The vertical dashed lines 
represent the dates of state transitions. The arrows on top of the plot show 
the dates where QPO are detected and to which type they belong (see paper 1).}
\label{fig:hardstate}
\end{figure}

\begin{figure}
\plotone{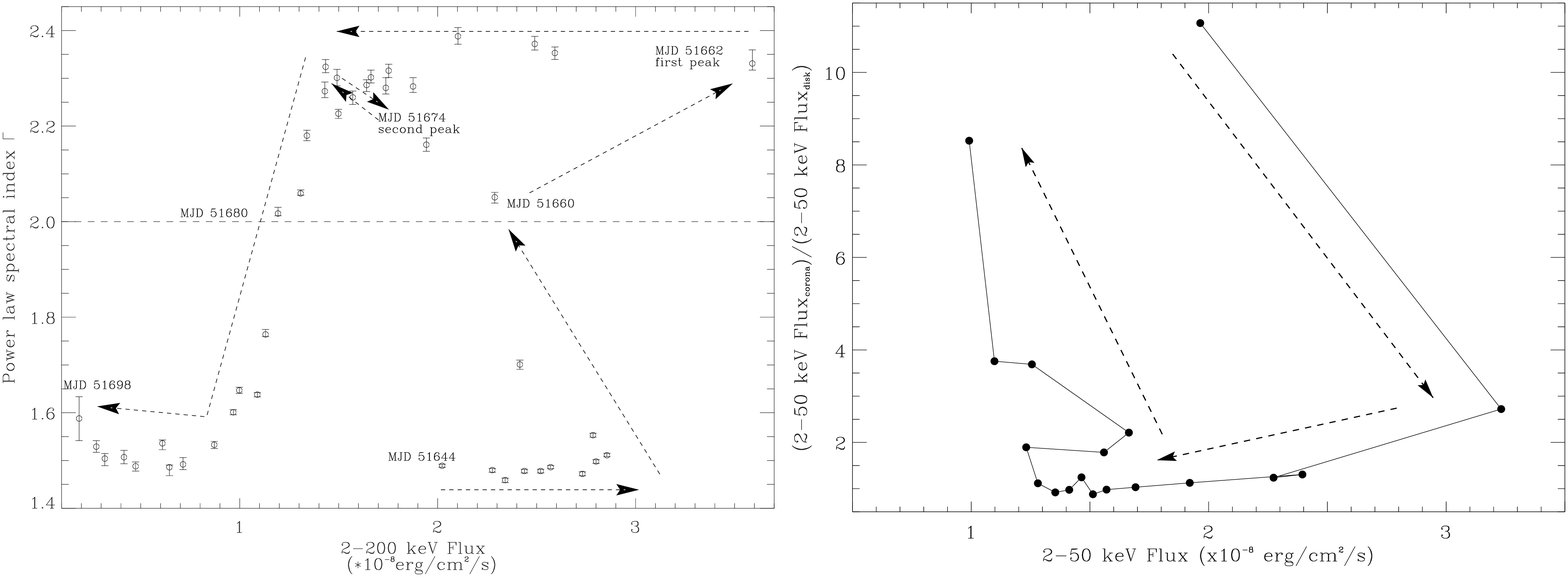}
\caption{{\bf{Left :}} Evolution of the power law photon index $\Gamma$ vs. 
the 2-200 keV flux for the whole period of outburst. The hysteresis discussed in the text is clearly visible on this figure. The horizontal line separates the two 
states, and the arrows indicate the chronology of the events. Some dates, 
showing special events, have been reported. {\bf{Right :}} Ratio of the fluxes of
the powerlaw component (corona) to the multi-color disk component versus the source flux.
The errors are left unplotted for the sake of clarity. This plot allows for a direct comparison with the neutron star system Aql X--1 (Maccarone \& Coppi 2003). The arrows indicate the chronology of the outburst.}
\label{fig:alphavsflux}
\end{figure}

\begin{figure}
\plotone{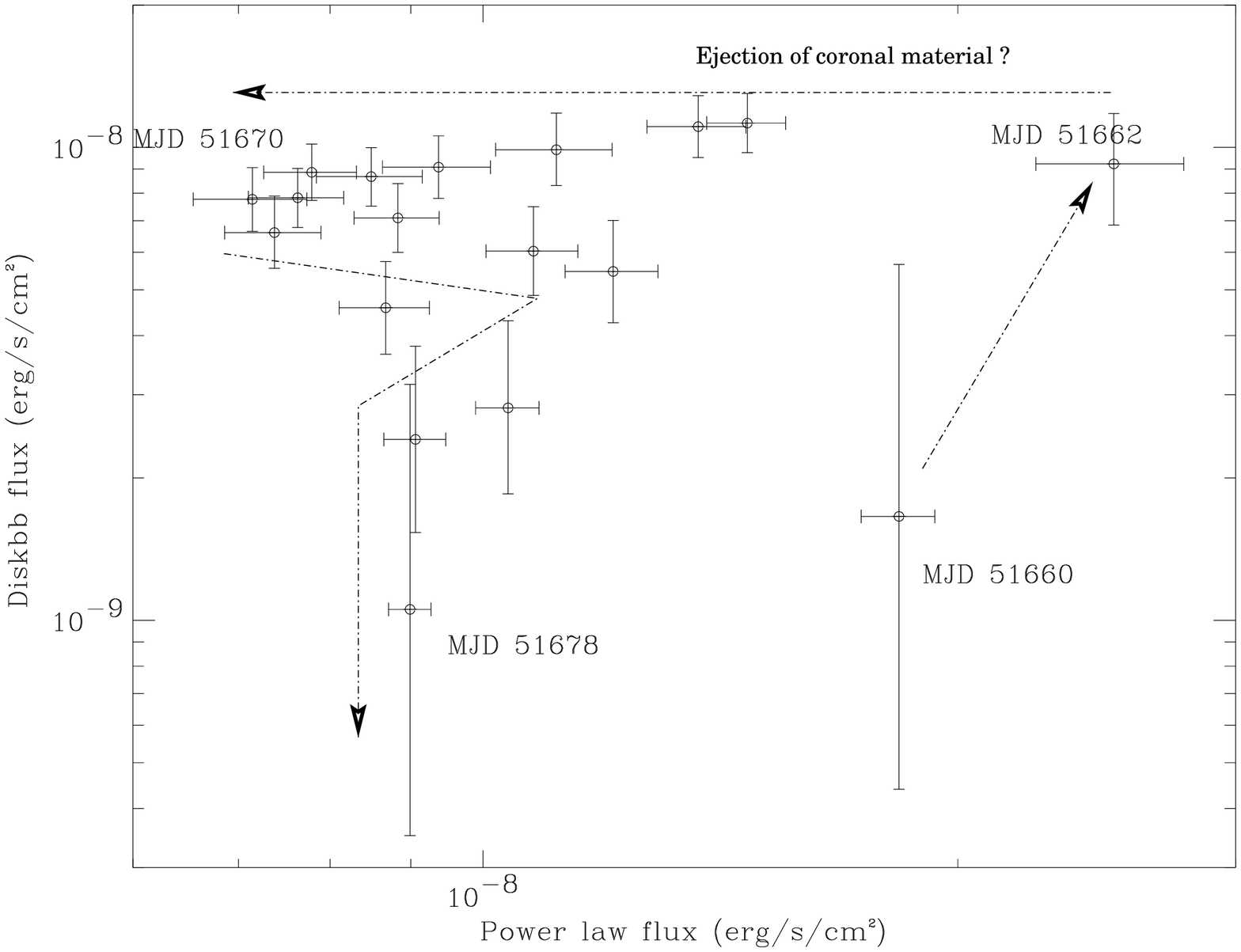}
\caption{Evolution of the disk blackbody 2--50 keV flux vs. the power-law 
2--50 keV flux. The arrows indicate the chronology of the events starting
 at the first transition on MJD 51660, and lasting on MJD 51678. Due 
to high uncertainties on the disk parameters, the points corresponding to 
MJD 51658, 51680 and 51682 are not represented here for the sake of clarity.}
\label{fig:Fluxbb}
\end{figure}

Corbel et al. (2001) report the detection of radio emission with a flux density 
$\propto\nu^{-0.46}$. The negative spectral index is indicative of optically thin 
synchrotron emission, that Corbel et al. (2001) attribute to a discrete ejection of material 
from the system. This detection of a discrete ejection, which might be associated with the state 
transition (Corbel \etal 2001), raises the question of the origin of the 
ejected material.
 If the ejection is associated with the transition, this may indicate that 
the Comptonizing medium is the source of the ejection of the material as suggested 
in \G1915 (Rodriguez \etal 2002b). The ejection may also be triggered at 
the peak of luminosity, soon after MJD 51662. 
Indeed, if we plot the 2--50 keV unabsorbed black body flux vs. the 2--50 keV 
unabsorbed power-law 
flux (Fig. \ref{fig:Fluxbb}), it appears that after the $1$ Crab flare (the
 extreme right point in Fig. \ref{fig:Fluxbb}), as the total luminosity
 is decreasing (Fig.\ref{fig:hardstate}), the disk blackbody flux remains  
$\sim$ constant during an initial period, while it is the power-law flux which 
decreases significantly. Given
 the relative constancy of the photon index those days (Fig. 
\ref{fig:hardstate} and \ref{fig:alphavsflux}), this may suggest that 
 there is less Compton up-scattering of the soft photons. A possible reason is 
that a part of the Compton medium is ejected. Given 
the detection of ejected material, it is tempting to consider that  
the Compton medium may be blown away, and not, as usually assumed, the 
innermost part of the disk.\\
\indent Here again it is interesting to compare this outburst with the 1998-1999 
one. During the latter outburst a discrete ejection of relativistic plasma (with apparent 
superluminal motion) was also detected right after the X-ray peak luminosity (Hannikainen et al. 
2001). Both outbursts have very similar initial stages, starting with LS, reaching a 
peak in luminosity soon after a state transition, and a discrete ejection coincident 
with the spike in the 1998 outburst, and possibly coincident with the spike during
the 2000 outburst. It is very interesting to note the similarities
in the behavior of the source during both outbursts. In 1998 the huge
X-ray spike is associated with large superluminal ejection (Hannikainen et al. 2001,
Corbel et al. 2002), during the 2000 outburst the X-ray spike is associated 
with a discrete ejection of smaller amplitude. This may point out a strong coupling
between the X-ray behavior and ejection of material in \X1550.\\
\indent However the observation of Corbel \etal (2001) does not allow us to determine
 the date of ejection. 
Future multi-wavelength monitoring of SXTs in outburst, should
 allow to study better the accretion-ejection coupling as has been done
 for jet sources e.g. \G1915, GX 339--4.
In the near future, this can be done also by including in multi-wavelength studies 
the high energy broad band spectra from the soft X-rays with RXTE, up to the 
gamma rays with INTEGRAL.

\begin{acknowledgements}
J.R. would like to thank  D. Barret, Ph. Durouchoux, J.-M. Hameury, G. Henry, 
E. Kalemci, M. Tagger, and P. Varnière for useful discussions. J.R. 
acknowledges financial support from the French spatial agency (CNES). 
J.A.T. acknowledges partial support from NASA grant NAG5-13055.
The authors thanks the anonymous referee for useful comments.
This research has made use of data obtained through the High Energy 
Astrophysics Science Archive Center Online Service, provided by the 
NASA/Goddard Space Flight Center.
\end{acknowledgements}

\newpage

\begin{thebibliography}{}

\bibitem{bell01} Belloni T., 2001, proceedings of the JHU/LHEA workshop, ``X-rays from Accretion onto Black Holes'', June 20-23 / astro-ph 0112217.

\bibitem{cor01} Corbel S., Kaaret P., Jain R.K., Baylin C.D., Fender R.P., Tomsick J.A., Kalemci E., McIntyre V., Campbell-Wilson D., Miller J.M., McCollough M.L., 2001, ApJ, 554, 43.

\bibitem{Corbel02a} Corbel S., Fender R.P., Tzioumis A.K., Tomsick J.A., Orosz J.A., Miller J.M., Wijnands R., Kaaret P., 2002, Science, 298, 196.

\bibitem{corbel02b} Corbel S. \& Fender R.P., 2002, ApJ, 573, 35.

\bibitem{Frank92} Frank J., King A., Raine D., 1992, ``Accretion Power in Astrophysics'', Cambridge Astrophysics Series.

\bibitem{hani01} Hannikainen D., Campbell-Wilson D., Hunstead R., McIntyre V., Lovell J., Reynolds J., Tzioumis T., Wu K., 2001, Ap\& SS, 276, 45.

\bibitem{hom01} Homan J., Wijnands R., van der Klis M., Belloni T., van Paradijs J., Klein-Wolt M., Fender R.P., M\'endez M., 2001,ApJS, 132,377.

\bibitem{jain01} Jain R.K., Baylin C.D., Orosz J.A., McClintock J.E., Remillard R.A., 2001, ApJ, 554, L181.

\bibitem{kaaret02} Kaaret P., Corbel S., Tomsick J.A., Fender R., Miller J.M., Orosz J.A., Tzioumis T., Wijnands R., 2003, ApJ, 582, 945.

\bibitem{kong02} Kong A.K.H., McClintock J.E., Garcia M.R., Murray S.S.,  Barret D., 2002, ApJ, 570, 277.

\bibitem{kal01} Kalemci E., Tomsick, J. A., Rothschild R. E., Pottschmidt K., Kaaret P.,2001, ApJ, 563, 239 , {\bf{K01}}.

\bibitem{maccarone02} Maccarone T.J., Coppi P.S., 2003, MNRAS, 338, 189.

\bibitem{markoff01} Markoff S., Falcke H., Fender, R., 2001, A\&A, 372, L25.

\bibitem{markoff03} Markoff S., Nowak M.A., Corbel S., Fender R.P., Falcke H., 2003, A\&A, 397, 645.

\bibitem{mendez96} Méndez M., van der Klis M., 1997, ApJ, 479,926.

\bibitem{MFR99} Merloni A., Fabian A. C., and Ross R.R., 2000,  MNRAS, 313, 193.

\bibitem{mil01} Miller J.M., Wijnands R., Homan J., Belloni T., Pooley D., Corbel S., Kouveliotou C., van der Klis M., Lewin H.G., 2001, ApJ, 563, 928.

\bibitem{mil03} Miller J.M., Marshall H.L., Wijnands R., DiMatteo T., et al., 2003, MNRAS 338, 7.

\bibitem{Mitsuda84} Mitsuda K., Inoue H., Koyama K., Makishima K., Matsuoka M., Ogawara Y., Suzuki K., Tanaka Y., Shibazaki N., Hirano T., 1984, PASJ, 36, 741.

\bibitem{Miyamoto95} Miyamoto , S., Kitamoto S., Hayashida K., Egoshi W., 1995, ApJ, 442, L13.

\bibitem{Nowak02} Nowak M.A., Wilms J., Dove J.B., 2002, MNRAS, 332, 856.

\bibitem{Orosz01} Orosz J.A., Groot P.J., van der Klis M., McClintock J.E., Garcia M.R., Zhao P., Jain R.K., Bailyn C.D., Remillard R.A., 2002, ApJ, 562, 568.

\bibitem{Reilly01} Reilly K.T., Bloom E.D., Focke W., Giebels B., Godfrey G., Saz Parkinson P.M., Shabad G., Ray P.S., Bandyopadhyay R.M., Wood K.S., Wolff M.T., Fritz G.G., Hertz P., Kowalski M.P., Lovelette M.N., Yentis D.J., 2001, ApJ, 561, L183.

\bibitem{Rem99}  Remillard R.A., McClintock J.E., Sobczak G.J., Bailyn C.D.,
 Orosz J.A., Morgan E.H., Levine A.M., 1999, ApJ, 517, L130.

\bibitem{Remillard02b} Remillard R.A., Muno M.P., McClintock J.E., Orosz J.A., 2002,  ApJ, 580, 1030.

\bibitem{Rodriguez02a} Rodriguez J., Varni\`ere P., Tagger M., Durouchoux P., 2002a, A\&A, 387, 487.

\bibitem{Rodriguez02b} Rodriguez J., Durouchoux P., Mirabel F., Ueda Y., Tagger M., Yamaoka K., 2002b, A\&A, 386, 271.

\bibitem{paper1} Rodriguez J., Corbel S., Kalemci E., Tomsick J.A., Tagger M., 2002, {\em submitted to ApJ}, paper 1.

\bibitem{rutledge99} Rutledge R.E., Lewin W.H.G., van der Klis M., van Paradijs J., Dotani T., Vaughan B., Belloni T.,  Oosterbroek T., Kouveliotou C., 1999, ApJS, 124,265.

\bibitem{smith98} Smith D.A., 1998, IAU Circ. 7008.

\bibitem{smith00b} Smith D.A., Levine, A.M., Remillard R., Fox D., Schaefer R., RXTE/ASM Team, 2000, IAU Circ. 7394.

\bibitem{Smith02} Smith D.M., Heindl W.A., Swank J.H., 2002, ApJ, 569, 362.

\bibitem{sob99} Sobczak G.J., McClintock J.E., Remillard R.A., Levine A.M., Morgan E.H., Baylin C.D., Orosz J.A., 1999, ApJ, 517, L121.

\bibitem{sob00} Sobczak G.J., McClintock J.E., Remillard R.A., Cui W., Levine A.M., Morgan E.H., Orosz J.A., Baylin C.D., 2000, ApJ, 531, 537.

\bibitem{tom01} Tomsick J.A., Corbel S., Kaaret P., 2001, ApJ, 563, 229.

\bibitem{wilson01} Wilson C.D., Done C., 2001, MNRAS, 325, 167.

\bibitem{wu02} Wu K., Soria R., Campbell-Wilson D., Hannikainen D., Harmon B.A., Hunstead R., Johnston H., McCollough M., McIntyre V., 2002, ApJ, 565, 1161.

\bibitem{zhang97} Zhang S.N., Harmon B.A., Paciesas W.S., Remillard R.E., van Paradijs J.,  ApJ, 477, L95, 1997.

\end{thebibliography}
\end{document}